# Electronic properties of type-II Weyl semimetal $WTe_2$. A review perspective.

P. K. Das[1], D. Di Sante[2], F. Cilento[3], C. Bigi[4], D. Kopic[5], D. Soranzio[5], A. Sterzi[3], J. A. Krieger[6,7,8], I. Vobornik[9], J. Fujii[9], T. Okuda[10], V. N. Strocov[6], M. B. H. Breese[1,11], F. Parmigiani[3,5], G. Rossi[4,9], S. Picozzi[12], R. Thomale[2], G. Sangiovanni[2], R. J. Cava[13], and G. Panaccione[9,*]

[1]*Singapore Synchrotron Light Source, National University of Singapore, 5 Research Link, 117603, Singapore*

[2]*Institut für Theoretische Physik und Astrophysik, Universität Würzburg, Am Hubland Campus Süd, Würzburg 97074, Germany*

[3]*Elettra - Sincrotrone Trieste S.C.p.A., Strada Statale 14, km 163.5, Trieste 34149, Italy*

[4]*Dipartimento di Fisica, Universitá di Milano, Via Celoria 16, I-20133 Milano, Italy*

[5]*Universitá degli Studi di Trieste - Via A. Valerio 2, Trieste 34127, Italy*

[6]*Paul Scherrer Institute, Swiss Light Source, CH-5232 Villigen, Switzerland*

[7]*Laboratory for Muon Spin Spectroscopy, Paul Scherrer Institute, CH-5232 Villigen PSI, Switzerland*

[8]*Laboratorium für Festkörperphysik, ETH Zürich, CH-8093 Zürich, Switzerland*

[9]*Istituto Officina dei Materiali (IOM)-CNR, Laboratorio TASC, in Area Science Park, S.S.14, Km 163.5, I-34149 Trieste, Italy*

[10]*Hiroshima Synchrotron Radiation Center (HSRC), Hiroshima University, 2-313 Kagamiyama, Higashi-Hiroshima 739-0046, Japan.*

[11]*Department of Physics, National University of Singapore, 117576, Singapore*

[12]*Consiglio Nazionale delle Ricerche (CNR-SPIN), c/o Univ. Chieti-Pescara "G. D'Annunzio", 66100 Chieti, Italy*

[13]*Department of Chemistry, Princeton University, Princeton, New Jersey 08544, USA*

\* Corresponding author: giancarlo.panaccione@elettra.eu

*Currently, there is a flurry of research interest on materials with an unconventional electronic structure, and we have already seen significant progress in their understanding and engineering towards real-life applications. The interest erupted with the discovery of graphene and topological insulators in the previous decade. The electrons in graphene simulate massless Dirac Fermions with a linearly dispersing Dirac cone in their band structure, while in topological insulators, the electronic bands wind non-trivially in momentum space giving rise to gapless surface states and bulk bandgap. Weyl semimetals in condensed matter systems are the latest addition to this growing family of topological materials. Weyl Fermions are known in the context of high energy physics since almost the beginning of quantum mechanics. They apparently violate charge conservation rules, displaying the "chiral anomaly", with such remarkable properties recently theoretically predicted and experimentally verified to exist as low energy quasiparticle states in certain condensed matter systems. Not only are these new materials extremely important for our fundamental understanding of quantum phenomena, but also they exhibit completely different transport phenomena. For example, massless Fermions are susceptible to scattering from non-magnetic impurities. Dirac semimetals exhibit non-saturating extremely large magnetoresistance as a consequence of their robust electronic bands being protected by time reversal symmetry. These open up whole new possibilities for materials engineering and applications including quantum computing. In this review, we recapitulate some of the outstanding properties of $WTe_2$, namely, its non-saturating titanic magnetoresistance due to perfect electron and hole carrier balance up to a very high magnetic field observed for the very first time. It also hosts the Lorentz violating type-II Weyl Fermions in its bandstructure, again first predicted candidate to host such a remarkable phase. We primarily focus on the findings of our ARPES, spin-ARPES, and time-resolved ARPES studies complemented by first-principles calculations.*

### I. Introduction:

One of the most exciting discoveries in condensed matter physics in recent years, besides graphene and topological insulators, is the experimental realization of the Weyl Fermion state in condensed matter systems. The Weyl state was first predicted in the high energy physics context of weak interactions [1]. However, it has now been experimentally observed in condensed matter systems as a topological low energy quasiparticle excitation. The notion of topology in condensed matter physics dates back to the discovery of the quantum Hall effect [2, 3], where the Hall conductance is found to be quantized in units of $e^2/h$. The Quantum Hall state is fundamentally different from others of Landau's spontaneous symmetry breaking phases like crystallization, magnetism, superconductivity, etc. In the case of topological phases, the transition does not break any symmetry, and the phases are not associated with a local order parameter. Certain fundamental properties of a topological system change only through a quantum phase transition, and, as such, are not affected by continuous modifications in material's parameters until the phase transition occurs. Another remarkable example of topological phases in condensed matter systems is the discovery of topological insulators, where the combination of strong spin orbit coupling and time reversal symmetry results in a bulk bandgap similar to that of an ordinary insulator and a protected metallic surface electronic structure [4-9].

Among 2D systems, Transition Metal Dichalcogenides (TMDs) display outstanding electrical, mechanical and optical properties. TMDs are a group of materials with chemical formula $MX_2$, where M is a transition metal (from group 4-10; for example, Ti, Zr, V, Nb, Ta, Mo, Te, W etc.), and X is a chalcogen atom (S, Se, or Te). Generally, they form a layered structure where a layer of metal (M) atoms is sandwiched between two chalcogen layers; these dichalcogenide layers are then stacked along a particular direction to form the bulk structure. The atomic bonding within the layers shows a strong covalent nature, while the layers themselves are weakly held together via van der Waals attractions - that is why the TMDs are commonly believed to be structurally and electronically two dimensional. Moreover, although TMDs present mainly a layered structure (only some TMDs of group 8–10 are generally found in non-layered structures), the presence of two constituent chemical species results in a predominant semiconducting nature, in contrast to semimetallic graphene, placing TMDs in a favorable perspective for electronic and spintronics applications. Their peculiar metal-semiconductor crossover as a function of thickness, and their semimetallic character, recently revamped interest in the magnetic field response of TMDs, with prototypical examples found in $WTe_2$ and $NbSe_2$ [10-12]. In general, TMDs have been studied extensively in the past few decades due to their wide range of physical, chemical, electronic, optical and thermal properties [13-15]. Depending on their crystallographic structure and particular chemical combination, they exhibit unique electronic and optical properties ranging from metallic, semimetallic, and semiconducting phases to displaying various low temperature phenomena like superconductivity [16] and charge density waves [17]. Moreover, they have been proposed for various real-life applications including transistors, photodetectors, and electroluminescent devices [18]. Even though the layers are weakly coupled in the TMDs, their electronic and physical properties highly depend on the number of layers present. For example, bulk $MoS_2$ is an indirect bandgap (1.3 eV) semiconductor, while it turns into a direct bandgap system in the monolayer configuration with an increased bandgap of 1.8 eV [19, 20]. Reduction in the number of layers results in quantum confinement, and also changes the hybridization between the $p_z$ orbital of S atoms and $d$ orbitals of Mo atoms, ultimately causing drastic changes in the electronic structure. It is found from first-principles calculations that the band structure near the Brillouin zone $K$ point mainly consists of localized $d$ orbitals, which remains substantially unaffected when the number of layers is reduced. On the contrary, the electronic states near the Γ point are the result of $p_z - d$ hybridization and, as such, experience a significant influence of interlayer coupling [20]. Having a direct bandgap, monolayer MoS₂ turns out to be a suitable template for photoluminescence applications. Furthermore, it also exhibits valley polarization [21-23].

The layer dependence plays an important role in the case of topological insulators as well. The formation of topological states in Bi₂Se₃ requires six quintuple layers, and also the spin texture of the topological surface state evolves as a function of the number of layers [24, 25] up to several layers from the surface. The layer dependence is of pivotal importance in multilayer graphene, TMDs, and other related low-dimensional materials, which show several different bulk properties and can be engineered for various potential technological applications.

In the present work, we focus our attention on WTe₂, and review two of its peculiar characteristics: (1) the system exhibits extremely large non-saturating magnetoresistance (MR), and (2) it hosts type-II Weyl Fermions as low energy excitations. WTe₂ is a special member of the TMD family. There is an additional structural anisotropy in WTe₂ apart from its layered geometry - the X-M-X chains form a zig-zag arrangement along the crystallographic $a-$ direction, imposing on the material an effective 1D dimensionality. The observed MR in WTe₂ is indeed highly anisotropic. The largest MR effect is accomplished when the current flows along the Te-W-Te chains i.e., precisely along the crystallographic $a-$direction, and the field is applied perpendicularly to the layer stacking i.e., along the crystallographic $c-$direction. The MR effect drops by more than 90% when the magnetic field is applied along the other principle crystallographic directions.

The extremely large non-saturating MR takes its origin from the perfect compensation of electron and hole carriers in this material [10, 26]. The MR in WTe₂ can reach an astonishing 13 million per cent at 0.53 K at an applied magnetic field of 60 T without any tendency to saturate [10, 27]. Materials with large MR have several important technological applications, which include magnetic hard drives [28] and magnetic sensors [29, 30]. The MR observed in ordinary metals is relatively weak in magnitude, generally of the order of only few percent. Giant MR [28] and colossal MR [31] is observed in magnetic thin films, and manganese based perovskites. Magnetic materials generally exhibit negative MR, because the application of a magnetic field to a magnetic system brings more order in the system and consequently reduces the spin disorder scattering of the carriers. Extremely large MR has been observed in semimetallic bismuth [32] and graphite [33]. The large MR effect in semimetals is attributed to the resonance of electron and hole carrier balance, and the behavior is well explained by a two band model. However, the resonance condition in both graphite and bismuth is only guaranteed in the lower applied magnetic field regime. With an increase of the magnetic field intensity, the system falls off resonance, and eventually the MR saturates at higher field. It has also been recently discovered that the 3D Dirac semimetals $Cd_3As_2$ [34], $Na_3Bi$ [35], and the topological Weyl semimetals NbP [36], NbAs [37], and TaAs [38-40] are characterized by a non-saturating MR. In the case of topological semimetals, the MR is found to be linear with the applied magnetic field, where the electrons are protected against the backscattering because of time reversal symmetry. As the magnetic field is applied, the backscattering protection is progressively lifted up, and therefore the MR is expected to rise linearly.

Furthermore, it has been theoretically predicted [41] and experimentally verified [42-45] that $WTe_2$ hosts type-II Weyl Fermionic states. The type-II Weyl points exist at the frontier between electron and hole pockets in $WTe_2$, and result in surface Fermi arcs with extremely high electronic mobility. The Weyl points in $WTe_2$ are located above the chemical potential, making it impossible to detect them by equilibrium ARPES measurements. However, equilibrium ARPES can be used to keep track of the nontrivial Fermi arcs connecting the electron and hole pockets on the (001) surface. The magneto-transport anomalies in $WTe_2$ may also be related to its Weyl physics [46, 47]. Apart from these two main properties, $WTe_2$ also exhibits sizable thermoelectric power [48, 49], a temperature induced Lifshitz transition [49], and a pressure-induced transition to superconductivity with an associated change in the sign of the Hall coefficient [50]. Both the Lifshitz transition and the appearance of superconductivity are believed to be intertwined and resulting from a significant reconstruction of the Fermi surface topology driven by a quantum phase transition. Therefore, $WTe_2$ turns out to be a rich platform to study various novel quantum phenomena, and a candidate functional material for various devices application.

In this review, we have combined spin-, time- and angle-resolved photoemission spectroscopy, as well as first-principles calculations to shed light on the main properties of $WTe_2$, a system with extraordinary fundamental and technological implications. ARPES measurements allow us to verify that the electron and hole pockets are of comparable sizes. We then established the role of the layer dependence in the formation of such pockets from realistic ab initio supercell calculations. Spin-resolved ARPES studies, moreover, revealed a large spin polarization in this material, suggesting that strong spin orbit coupling effects play a leading role. Despite the layered structure, we reported a bulk modulation of the electronic properties from $k_z$ dependent soft-X-ray ARPES experiments. We have also performed time-resolved ARPES (TR-ARPES) experiments to reveal the out-of-equilibrium dynamics of the electronic bands near the Fermi energy. Finally, we discuss the possible role that electronic correlations, owing to the $d$ nature of W orbitals, might play in explaining the actual band structure of WTe$_2$.

## II. Experiment:

High quality single crystalline samples of WTe$_2$ are used in all of our experiments. The single crystals were grown by Bromine vapor transport method, and are formed in platelet forms with typical dimension 0.1 mm × 2 mm × 5 mm, with crystallographic c-direction perpendicular to the planar surface. The electronic band structure was measured at the CNR-IOM APE beamline [51] by the ARPES method. The beamline is equipped with a Scienta DA30 electron analyzer and VLEED spin detectors for spin resolved study [52]. The principle of the ARPES technique is based on the photoelectric effect [53]. A material surface is illuminated by the light of a known photon energy (and also of known polarization when a synchrotron or laser source is used). The energy of the photon must be higher than the work function of the material in order to photoemit electrons from the metal surface. The photoemitted electrons are collected and their kinetic energy is determined by an electron analyzer. The emission angle of the photoemitted electrons is also recorded. The energy of photoelectrons and photoemission angle enable us to determine the momentum of the photo electrons because in the process of photoemission, the energy and momentum of electrons are conserved. Since the momentum component parallel to the surface is conserved in the photoemission process, it allows us to map the momentum component of the electronic bands parallel to the surface. In the actual experiment, a single crystalline sample is required with an atomically flat surface for a well-defined momentum vector parallel to the surface of the sample. In our present case, a high quality single crystalline sample of WTe$_2$ was cleaved in-situ by using Kapton tape or using top-post method. The crystallographic orientation and surface quality of the samples were verified by the Low Energy Electron Diffraction (LEED) pattern. The band structure measurements were performed at liquid nitrogen temperature (77 K), and various different photon energies were used. In case of spin-resolved ARPES, in addition to the energy and momentum of the electrons, the spin of the electrons is also determined. The spin detector used in our measurement utilizes highly efficient VLEED based spin polarimeter [52]. The photoemitted electrons are channeled to a pre-oxidized magnetic target using electrostatic lenses. The intensity of the reflected electronic counts depends on the magnetization direction of the target and on the spin polarization of the incoming photoemitted electrons. Measuring the reflected electron beam with two opposite orientations of the target magnetization allows us quantification of spin asymmetry of the electronic bands. The spin asymmetry is given by $A = (I_+ - I_-)/(I_+ + I_-)$, where $I_+$ and $I_-$ are reflected VLEED electronic counts corresponding to positive and negative target magnetizations, respectively. Once the spectra is compared with a known spin polarization material, the actual spin polarization of the unknown system can be determined [52]. The spin polarization is quantified as $P = A/S_{eff}$, where $S_{eff}$ is the effective Sherman function, which depends on the detector and instrument setup, and determined by comparing polarization of a known system (e.g., the VLEED setup at APE, Elettra is calibrated using fully polarized Rashba split bands of Au (111) surface states as reference). Finally, the up and down spin component is calculated by the following expression: $S_\uparrow (S_\downarrow) = (1 \pm P)(I_+ + I_-)/2$. The spin resolved measurements presented here were performed at BL-9B of Hiroshima Synchrotron Radiation Center (HiSOR), Japan. The beamline utilizes Scienta R4000 electron analyzer, and a similar VLEED setup [54, 55] as described above. The VLEED setup was calibrated using spin polarized bands of Bi (111) surface, and a $S_{eff}$ value of 0.2 was obtained. The bulk sensitive soft-x-ray ARPES measurements were performed at the ADRESS beamline [56, 57] at the Swiss Light Source, Switzerland. Photon energies from 400-800 eV were used to map the band structure perpendicular to the direction of the stacking layers over about nine Brillouin zones. Time-resolved ARPES measurements are carried out at T-REX beamline at FERMI/ELETTRA Sincrotrone, Trieste.

**First-Principles calculations:**

Density functional theory based calculations were performed using the VASP ab-initio simulation package [58, 59]. Consistent with previous theoretical reports, the local density approximation within the PBE parametrization for the exchange-correlation potential was used, by expanding the Kohn-Sham wavefunction into plane-waves up to an energy cut-off of 400 eV [60]. The Brillouin zone was sampled on a 24×12×4 regular mesh. Spin-orbit coupling (SOC) was included in a self-consistent manner. Moreover, to go beyond the single particle picture, we used the LDA+U to treat electronic correlations within an energy independent Hartree description. Atomic positions of WTe$_2$ were fully relaxed starting from experimental data in Ref. [61].

### III. Results and discussions:
#### a. Crystal structure:
WTe$_2$ crystallizes in an orthorhombic crystal structure with space group $Pmn2_1$. The lattice parameters are $a = 3.477$ Å, $b = 6.249$ Å, and $c = 14.018$ Å such that 4 formula units are incorporated in the unit cell. The $Pnm2_1$ space group is noncentrosymmetric; it has a mirror plane symmetry in its $bc$ plane, and a glide plane symmetry in the $ac$ plane followed by a (0.5, 0, 0.5) translation. It is important to note that the noncentrosymmetric structure is a necessary requirement for the existence of Weyl fermions in a nonmagnetic system like WTe$_2$ [41]. In addition, the absence of inversion symmetry gives rise to the possibility of the lifting up of spin degeneracy in the electronic structure. The W layer is sandwiched between two Te layers, and strongly linked by covalent bonds. These Te-W-Te layers are then stacked along the crystallographic $c$-direction, and held together via weak van-der Waals interactions. The crystal structure of WTe$_2$ is shown in Fig. 1a. In the specific case of WTe$_2$, the atoms form a zig-zag chain (Te-W-Te) along the crystallographic $a$-direction, effectively creating an one dimensional substructure within the layers.

**b. ARPES observations:**
We begin with the observed electronic structure by high resolution ARPES. Fig. 1b presents the band dispersion measured along the $X - \Gamma - X$ direction using a photon probe energy of 68 eV and at temperature of 77 K. The main feature of interest, i.e. the existence of electron and hole pockets at the Fermi energy, the key behind the titanic magnetoresistance of WTe$_2$, is clearly resolved and in agreement with the other APRES reports [62, 63]. Based on the observation of quantum oscillations [64] and first-principles calculations, there is an almost overlapping pair of electron and hole pockets at both sides of the Brillouin zone center (eight pockets in total). Each of the smaller pockets are completely inside the larger pockets. From the observed spectral weight, we cannot exclude the possibility of a small hole pocket at the Brillouin zone center, as reported in other studies [63]. The areas of the electron and hole pockets are equal within our experimental error bar. The perfect electron and hole balance persists up to very high magnetic fields, unlike any other previously known material. In the case of bismuth, for instance, the MR saturates at about 40 T magnetic field [65]. WTe$_2$ is the first known system to show a perfectly balanced electron and hole concentration, and almost quadratic MR up to a field as high as 60 T [10] without any tendency of saturation. The closeness between the two pockets and the small overlap between the conduction band minima and valence band maxima might have played important role in the extreme MR of this system [62]. From the Fermi surface map in Fig. 1d, it is seen that WTe$_2$ possesses a highly anisotropic electronic structure, which is reflected in its highly anisotropic magnetotransport properties. The Fermi surface is remarkably unidirectional; extended along the $X - \Gamma - X$ direction while highly restricted along the perpendicular $Y - \Gamma - Y$ direction. The scattering of electron and holes would inherit this unidirectionality; the electron and holes will prefer to flow along the $X - \Gamma - X$ direction i.e., along the direction of Te-W-Te chains.

## c. Magnetoresistance and two-band model:

From the semi-classical theory of MR, it can be shown that in order to obtain non-saturating behavior, perfect electron and hole balance is a necessary condition. The MR in WTe$_2$ increases quadratically with the applied field up to a very high magnetic field. The quadratic MR behavior is reminiscent of the perfect balance of electron and hole compensation in the material. According to the semiclassical two-band model, the conductivity tensor ($\hat{\sigma}$) is given by the following expression:

$$\hat{\sigma} = e\left[\frac{n\mu_e}{1+i\mu_e B} + \frac{p\mu_h}{1+i\mu_h B}\right] \qquad (1)$$

where $n$ and $p$ are the electron and hole carrier densities, and $\mu_e$ and $\mu_h$ are the carrier mobilities of electrons and holes, respectively. The inverse of the conductivity tensor gives the resistivity tensor ($\hat{\rho}$). The longitudinal resistivity and the MR are then given by the following expressions:

$$\rho_{xx}(B) = Re(\hat{\rho}) = \frac{1}{e} \frac{(n\mu_e + p\mu_h) + (n\mu_h + p\mu_e)\mu_e\mu_h B^2}{(n\mu_e + p\mu_h)^2 + (n-p)^2 \mu_e^2 \mu_h^2 B^2} \qquad (2)$$

$$MR = \frac{\sigma\sigma'\left(\frac{\sigma}{n} + \frac{\sigma'}{p}\right)^2 \left(\frac{B}{e}\right)^2}{(\sigma+\sigma')^2 + \sigma^2 \sigma'^2 \left(\frac{1}{n} - \frac{1}{p}\right)^2 \left(\frac{B}{e}\right)^2} \qquad (3)$$

with $\sigma$ and $\sigma'$ being the electrical conductivities of electrons and holes, respectively. From the above equation, in the $n \to p$ limit, we find that the MR is proportional to the square of the magnetic field,

$$MR = \frac{\sigma\sigma'\left(\frac{B}{e}\right)^2}{n^2} = \mu_e\mu_h B^2 \qquad (4)$$

The observation of nearly equal electron and hole pocket area qualitatively explains therefore the quadratic increase of MR as a function of the external magnetic field.

Since the MR effect of WTe$_2$ depends on the precise carriers balance, it is therefore highly dependent on the Fermi surface topography, and affected by the small details of the electronic structure. Such small changes in the Fermi surface can be induced by temperature, pressure, chemical doping or electrical gating, resulting in drastic changes of its MR behavior. It is worthwhile to mention that another type of non-saturating MR has been recently observed in Dirac semimetallic systems [34], where the MR is linear with the field. In the case of the Dirac semimetals, the backscattering of the carriers is prohibited and protected by the time reversal symmetry. As the magnetic field is applied, this protection is gradually lifted resulting in the rise of the electrical resistivity. The Weyl topological properties of $WTe_2$ might also be involved in this MR behavior.

### d. "Turn on" behavior and Lifshitz transition:

The MR effect in $WTe_2$ is significantly reduced above the so-called 'turn on' temperature of 150 K. This 'turn on' behavior is not associated with any charge density wave, structural transition, or other types of electronic instabilities which are commonly observed in other TMD compounds [10]. The 'turn on' temperature is found to shift towards higher temperature with increasing applied magnetic field, which indicates that a scattering mechanism is likely in play behind this phenomenon. A temperature dependent ARPES study reported a reduction in the size of the hole pocket and an increase in the size of the electron pocket with increasing temperature [62]. The imbalance from the perfect electron-hole resonance condition at elevated temperature could be the reason of this 'turn on' behavior. A temperature induced Lifshitz transition, i.e. a change in the Fermi surface topology, is also reported at 160 K [49], around the same temperature range of 'turn on' point. At the onset of the Lifshitz transition, the hole pockets completely disappear and the electron pockets expands in the Brillouin zone. This restructuring of the Fermi surface is caused by the shift of the chemical potential with the temperature. The same study also reported a change in the slope of the temperature dependent thermoelectric power about the same temperature. At lower temperature, the MR scales with the Kohler's exponent of 1.98, very close to the quadratic behavior. However, as the temperature is raised the Kohler's scaling breaks down in the temperature range 70 – 140 K [49, 66]. These evidences further corroborate that the MR related phenomenon crucially depends on the precise balance of the electron-hole resonance and on the tiny details of the Fermi surface. It has been reported from temperature dependent Hall measurements that the hole carrier density suddenly increases below 160 K. The electron density is found to have a significant reduction below 50 K, leading to a nearly perfect electron and hole resonance at low temperature [67].

Likewise to the temperature induced Lifshitz transition discussed above, a pressure induced Lifshitz transition is also reported in $WTe_2$ [50], accompanied by the suppression of the MR effect and the appearance of superconductivity. This transition also occurs in the absence of any associated structural phase transition. With increasing applied pressure, the MR effect is found to be gradually suppressed and eventually completely disappears at 10.5 GPa, and the superconductivity arises exactly at the same pressure with a critical temperature $T_c$ = 2.8 K. The superconducting temperature reaches a respectable value of 6.5 K at 13.0 GPa, after that the $T_c$ decreases with increasing pressure. When a higher pressure is applied, there is a significant reduction in the lattice constant $c$ without exhibiting any structural phase transition. On the other hand, the in-plane lattice constants are not significantly affected by the application of external pressure. Electronically, since the *5p* orbitals of Te and *5d* orbitals of W are spatially extended, the band structure is prone to be very sensitive to any change to the lattice constants, and hence to the external pressure. Therefore, the Fermi surface reconstruction could be related to the anisotropic transformation of the Bravais lattice upon change of pressure. From the high-pressure Hall effect measurement, it is found that the density of hole carriers decreases with application of pressure while the carrier density of electrons increases, similar to what is observed in the case of the temperature driven Lifshitz transition. The Hall coefficient in this material is positive at ambient pressure, and gradually decreases with the applied external pressure. A change of the sign of the Hall coefficient is observed at the critical pressure of 10.5 GPa, corroborating the idea of a quantum phase transition associated with the drastic change in Fermi surface topology. Such a quantum phase transition is also in agreement with the observed Shubnikov-de-Haas quantum oscillations [68]. The pressure induced superconductivity is also verified by transport measurements [69], where the pressure is applied without any pressure-transmitting medium.

**e. Role of SOC and DFT results:**
In order to better understand the observed electronic structure, we have calculated the electronic structure of WTe$_2$ by first-principles calculations. The analysis is performed by assuming WTe$_2$ as pure 2D material, i.e. by looking at the bulk band structure for $k_z = 0$. We see that our DFT bands well reproduce the electron and hole pockets and other details of the band dispersion (Fig. 2 (a)). However, a closer inspection into the calculated bands reveals some differences with the observed spectra. For instance, the momentum positions of electron and hole pockets in the calculated bands are significantly offset from the Brillouin zone center compared to our experimental spectra. In addition, the maximum binding energy of the electron pocket is larger than the observed spectra by more than a factor of two. Next, we consider the importance of SOC in the better agreement between the experimental and theoretical spectra. WTe$_2$ consists of two heavy elements, i.e. tungsten and tellurium, therefore strong influence of SOC is expected in the formation of the band structure of the system. In Fig. 2(b) we show the calculated bandstructure upon the incorporation of SOC as superimposed over the experimental spectra. It is evident that it results in a remarkable better agreement between the electron and hole pocket locations in the Brillouin zone. The changes in the theoretical bandstructure with incorporation of SOC highlights the fact that relativistic effects are significant in this system.

Despite the fact that we obtain a better agreement between the calculated and experimental spectra after the inclusion of SOC, a quantitative agreement is not achieved yet. It is important to note here that there exists a non-negligible band dispersion in the direction perpendicular to the layers i.e., along the $k_z$ direction. However, the band dispersion along the $X - \Gamma - X$ direction cannot be fully reproduced by considering any specific $k_z$ value. It is also noted that even though the bulk calculation reproduces the details of various Fermi surface features, the calculated Fermi surface turns out to be always much larger than the experimental one [70]. This suggests that constraining ourselves to bulk calculations is not sufficient to explain the electronic properties of this system. We also note that the experimentally observed low energy final states of TMDs may differ from the free-electron dispersions, exhibiting non-parabolic dispersions and incorporating number of plane waves with different $k_z$ values [71].

**f. Layer dependence:**
Next, we discuss the role of dimensionality in the formation of perfect electron and hole balance in this system. After the discovery of graphene and topological insulators (TIs), the important role of dimensionality in the formation of topological states in those systems is seen in a new light. For example, the complete formation of topological surface states in TI requires at least six quintuple layers [24]. In addition, the spin texture of the topological surface states evolves with the number of layers [25]. In particular, dimensionality is also very critical in TMD systems. We have already mentioned that bulk MoS$_2$ exhibits an indirect bandgap semiconductor behavior, while it becomes a direct bandgap semiconductor in the monolayer. Therefore, we can conjecture that these materials are not truly two-dimensional, nor are they traditional bulk systems; they are 'less than 3D' or in-between 2D and 3D.

In order to elevate the agreement between theory and experiment at the quantitative level, here we have adopted a more realistic modeling where we explicitly take into account the contribution of each individual layer and project them over the surface Brillouin zone. Our model is based on a supercell made by van der Waals-bonded WTe$_2$ planes stacked along the crystallographic *c*-direction. We have separated the individual contribution of each Te-W-Te plane from the large number of bands in the supercell calculations by projecting the electronic band structure onto the atoms belonging to a given plane only. It must be noted that the actual spectra measured by ARPES represents a weighted spectral intensity of various layers, with the intensity of the deeper layers attenuated exponentially because of the finite mean free path of the photoemitted electrons inside the material. Such a direct connection of the layer-resolved dispersions with photoemission holds only for purely 2D states, where the interference between the layers vanishes. In Fig. 2, we have superimposed the supercell bands over the experimental spectra. On panels 2 (c-e), we present the projected band dispersions corresponding to individual first, second, and third layer, respectively. In this way, we are able to disentangle the role of each individual layer in the formation of various band structure features. In the plots of the supercell electronic states, the size of the circles is proportional to the contribution of that particular plane to the final spectra; the bigger the circle, the more that particular Te-W-Te layer contributes to the final spectral feature. Now comparing the band projections from the different topmost layers, we see that the electron pocket which is located at 0.35 Å$^{-1}$ from the zone center Γ is found already from the top layer, indicating the electron pocket has a strong surface character. On the other hand, the hole pocket at 0.2 Å$^{-1}$ arises only when we consider the third layer from the top. This indicates that the hole pocket represents a more bulk-like feature. The important point to note here from our layer-resolved density functional theory calculations is that they not only provide for a better agreement with the observed spectra, but they also further suggest that there is a profound connection between the dimensionality and the formation of electron and hole pockets. The electron and hole balance is achieved only when at least three Te-W-Te layers are considered from the top surface, and the balance is maintained within the bulk. In accordance to the bulk calculations, there is a non-negligible $k_z$ dispersion of the electronic states in this layered material. We have probed such band dispersion along the direction perpendicular the stacking of the layers by soft-x-ray ARPES. This investigation solidifies our conclusion from the layer resolved DFT calculations that one needs more than three layers to recover the electron-hole balance in this system.

### g. Bulk character of the electronic structure:

In order to experimentally access the bulk electronic structure of layered WTe$_2$, we have performed soft-x-ray ARPES measurements. By changing the photon probe energy from 400 eV to 800 eV and exploiting the photon energy dependence of electrons' mean free path, i.e., the depth dependence, we are able to map the electronic structure along the direction perpendicular to the WTe$_2$ layers over nine Brillouin zones. The $k_z$ dependence of the Fermi surface can be mapped from the photon energy dependence spectra by using the following expression: $k_z = [\left(\frac{2m_e}{\hbar^2}\right)(E_k \cos^2\theta - V_0)]^{1/2} + \kappa_{ph}$ , where $V_0$ is inner potential of the system [53], and $\kappa_{ph}$ is momentum component of photon along the surface normal. Use of high energy soft-X-ray photon as ARPES probe is very crucial for the better resolution of $k_z$ dispersion. The increase of the photoelectron mean free path in soft-X-ray energy range translates, by the Heisenberg uncertainty principle, to reduced intrinsic uncertainty of $k_z$ allowing thereby more accurate 3D band mapping [72]. In order to obtain the better insight of the three dimensional fermi surface of WTe$_2$, both the

$k_x$ vs $k_y$ and $k_x$ vs $k_z$ Fermi surfaces are presented in Fig. 3. Panel (a) shows the $k_x$ vs $k_y$ Fermi surface and the iso-energy cuts at deeper binding energy values. The calculated $k_x$ vs $k_z$ is shown in panel (b), while the experimental panel (c) shows the $k_z$ vs. $k_x$ Fermi surface map obtained from soft-X-ray ARPES measurement. We observe a clear 3D evolution of the bandstructure. Several important features are worthy of further consideration.

The observed periodicity of the $k_z$ evolution matches twice that of the Brillouin zone, i.e., as if only half of the length of the lattice vector along the crystallographic *c*-direction were considered. Note that the WTe$_2$ unit cell consists of two non-equivalent Te-W-Te layers. Such a double periodicity is not uncommon in ARPES spectra. It has been already observed in other nonsymmorphic crystal structures such as CrO$_2$ [73], graphite [74], and in TMDs such as 2H-WSe$_2$ [75] and 2H-NbSe$_2$ [76] where the photoemission selection rules determines the final state symmetry. Because of truly free-electron character of high-energy final states, the nonsymmorphicity effects are particularly clear in soft-X-ray ARPES spectra.

Another crucial feature we infer from our soft-X-ray data, differing from more surface sensitive VUV-ARPES, is the absence of any quasiparticle weight at the Brillouin zone center. The extremal orbits estimated from the quantum oscillation study [42] matches well with our VUV-ARPES data, where the areas of the electron and hole pockets are nearly equal. However, the Fermi surface measured using soft-X-ray photons shows that the area of the hole pocket is sensibly larger than that of the electron pocket. Therefore, it is essential to invoke a pronounced $k_z$ dispersion of the electronic band structure to explain the perfect electron-hole compensation in this material. Moreover, this observation supports the magnetotransport results, which bear evidence of three dimensionality in order to explain the extremely large effect [77]. A detailed analysis of our soft-X-ray study is presented in Ref. [78].

### h. Spin-polarization:
The investigation of spin-polarization properties of a material is important for two main reasons. Firstly, it gives a quantitative measure of the effect of SOC present in the material. Secondly, materials with large spin polarization are desirable for various spintronics applications. A large spin polarization is reported for the semiconducting TMD WSe$_2$ [79] as a result of the local asymmetry of the Se-W-Se layers. Despite WSe$_2$ possesses a global centrosymmetric crystal structure, the alternating layers locally carry a net opposite dipole moment, and therefore, the local spin texture is modulated. In the specific case of WTe$_2$, the crystal structure does not possess a center of inversion. Spin polarized electronic bands are expected in a non-centrosymmetric system consisting of two heavy elements, foreseeing a strong SOC effect. We have also seen that SOC plays important role in explaining various band features in WTe$_2$ upon comparing with the theoretical results. Although the magnitude of the spin polarization depends on several material factors, e.g., orbital character, bandgap, crystalline electric field etc., it primarily relies on the strength of SOC. Therefore, it is natural to expect significant spin polarization in this system. We measured the spin resolved EDCs at a number of *k*-points in the Brillouin zone. The probed *k*-point locations are indicated in Fig. 4d by colored markers on the Fermi surface; edge of hole pockets on the opposite sides of the Brillouin zone center (with green circle and black triangle, respectively); close to the Brillouin zone center along the $\Gamma - X$ direction (blue square); and close to the Brillouin zone center but away from the $\Gamma - X$ line (red diamond). We observed a strong spin polarization of electronic bands up to 40% at a binding energy value of 0.55 eV ($S_y$ component near the edge of hole pocket in Fig. 4a). We also find non-negligible spin polarization near the Fermi energy (Fig. 4b), which is pivotal to any "device applications" viewpoint. We observe that along the $X - \Gamma - X$ line, the spin texture is oriented in the perpendicular direction i.e., $S_x = 0$, $S_y \neq 0$ and $S_z \neq 0$. In other words, the spins are perpendicular to the Te-W-Te chains. The spin vector has both in-plane and out-of-plane non-vanishing components, unlike conventional Rashba systems and topological insulators, where the spin texture is distinguished by an in-plane orientation. It is also seen that the spins are oppositely oriented in the *k*-space (i.e., spin orientation is reversed at $\boldsymbol{k}$ and $-\boldsymbol{k}$ points), respectively, confirming that the spin polarization has a non-magnetic origin. In Fig. 4a and e, the spin polarization is probed at the positive and negative $k_x$ values, respectively, close to the $X - \Gamma - X$ line, where the spin directs perpendicularly to the $X - \Gamma - X$ direction. We find that the $S_y$ component of the spin direction is indeed reversed at opposite sides of the $\Gamma$ point. The observed opposite spin orientation on the opposite sides of $\Gamma$ might provide a mechanism to protect the electronic backscattering at zero magnetic field as previously suggested [63]. When an external magnetic field is applied, such a protection mechanism is gradually invalidated, with the concomitant increase of the electrical resistivity reminiscent of the MR behavior of WTe$_2$. The spin texture has been also calculated from first principles, and notable qualitative agreement is seen [70]. The experimental resolution was not enough to reproduce the fine details of the actual spin texture, which is extremely complex and characterized by large fluctuations in both energy and momentum. Our spin resolved ARPES results clearly suggest that the SOC is indeed strong, and plays an active role in the electronic structure of WTe$_2$, as also seen in the CD-ARPES results [63].

**i. Electronic correlations:**

Notwithstanding the fairly delocalized nature of 5*d* W orbitals, the inclusion of a modest amount of Coulomb interaction *U* in the *ab initio* calculations results in a sizable change of the electronic dispersion in the proximity of the chemical potential. As a function of increasing *U*, a shift of the electron pocket towards lower momenta and a clear modification of the hole pocket are observed as a general trend (see Fig. 5 (a-b)). For *U* = 2 eV, i.e. the value that provides a nice comparison between the measured and calculated Fermi surfaces [78], the two pockets exhibit a linear touching ~50 meV above the Fermi energy along the reciprocal direction $k_x$. It is likely that such a correlation-driven trend leads to a change of the topological properties of WTe$_2$, since the appearance of a type-I Weyl crossing has been recently proposed as a hallmark of topological transitions in non-centrosymmetric topological insulators [80]. Moreover, the DFT+U optical conductivity, as shown in Fig. 5 (c), nicely reproduces the main features of the measured spectrum, including the peak around 60 meV (~500 cm$^{-1}$) missing in standard DFT calculations, as discussed in Ref. [81]. As yet another indication towards an improved description of the optical properties, the shoulder at 150 meV (1400 cm$^{-1}$) as well as the infrared peak at 20 meV (160 cm$^{-1}$) are better captured by DFT+U.

**j. Topological aspects of the band structure:**

The Dirac equation admits three different representations for the Dirac spinor, which in the context of condensed matter translates into three types of relativistic fermions: Dirac, Majorana, and Weyl fermions. These are found to manifest themselves in several condensed matter systems as low energy quasi particle excitations. It has been recently proposed and experimentally observed that the Weyl fermions can exhibit two distinct types of band dispersions. The type-I Weyl points are associated with a point-like Fermi surface and result from a linear crossing of two topologically protected bands at the Fermi level. The type-I Weyl Fermions have been observed in the TaAs family of compounds [36-40, 82-84]. On the contrary, the type-II Weyl points are observed at the frontiers between electron and hole pockets, and the electronic dispersion is characterized by strongly tilted Weyl cones. As a consequence, the Lorentz invariance turns out to be broken in the case of type-II Weyl states. The transport properties of type-II Weyl semimetals are found to be remarkably different from those of the type-I counterparts based upon the differences of their Fermi surfaces. SOC is critical for the actual distribution of Weyl points in WTe$_2$. In the absence of SOC, there is a total of 16 Weyl points in the Brillouin zone. Once the effect of SOC is considered, half of the Weyl points which were previously present in the $k_z = 0$ plane annihilates [41]. Furthermore, the Weyl physics of the system is highly dependent on its lattice constants. A slight change in the lattice constants results in the annihilation of the Weyl points. It has been reported from first principles calculations that the WTe$_2$ band structure shows eight Weyl points when considering the low temperature lattice parameters in the (001) plane [41], whereas all the Weyl points annihilates when the slightly larger room temperature lattice constants [45] are considered.

The Weyl points in WTe$_2$ are predicted to be situated about 50 meV above the Fermi energy [41], making it impossible to observe them by the equilibrium ARPES experiment. However, the signature of the topological Fermi arc connecting the bulk electron and hole pockets can be observed in the (001) surface projection (see Fig. 5 (a)). From our data, we are unable to resolve the Fermi arcs associated with the Weyl physics of this material. The ARPES studies performed by other groups have reported the signature of the Fermi arcs on the (001) surface [42-45]. It is moreover reported that the topological states of WTe$_2$ are not observed with ARPES on certain types of surfaces [42]. Although WTe$_2$ possesses two non-equivalent surface terminations, the reason for the observation of topological states only on some particular surfaces is not related to the surface termination itself, but it could be rather related to the strain induced to the surface upon cleavage. The topological states in WTe$_2$ are indeed highly sensitive to strain and pressure as theoretically suggested [41]. It is also found that irrespective of the presence of the Weyl points, surface states connecting the bulk electron and hole pockets are present. The differences in the surface states between the trivial and topologically non-trivial phases are minute, and hardly detectable in experiments. Therefore, it is argued that the detection of surface states alone is insufficient to conclude that WTe$_2$ belongs to the family of type-II Weyl semimetals [45]. Apart from WTe$_2$, type-II Weyl Fermions have been observed in MoTe$_2$ [85, 86] and LaAsGe [87].

**k. Dynamics of the carriers on the femtosecond time scale:**
Finally, we discuss the non-equilibrium dynamics of the electronic bands of this material near the Fermi energy as investigated by Time-Resolved ARPES (TR-ARPES). TR-ARPES experiments on WTe$_2$ crystals were performed to the aim of elucidating the evolution of the electronic bandstructure on a femtosecond timescale, after excitation of the sample with intense near-infrared laser pulses. In this pump-probe spectroscopy, an intense laser pulse (pump) deposits energy into the material and brings it in an out-of-equilibrium condition, while a delayed weaker pulse of appropriate photon energy (probe) is used for the ARPES experiment. ARPES maps are then acquired for different values of the pump-probe delay t. Hence, TR-ARPES provides a direct view on the non-equilibrium electronic band structure; in this context, it is used to reveal how the electronic bands close to the Fermi level are modified after absorption of the pump pulse. In the experiments reported here, the excitation is performed via a 1.55 eV (800 nm) laser beam, with an absorbed fluence of 80 $\mu J/cm^2$, while the ARPES probe is performed with a 6.2 eV (200 nm) laser beam, obtained by frequency-mixings processes. The overall time-resolution of the experiment is 120 fs; the energy resolution is of the order 50 meV. The temperature of the sample during the experiments is kept at T = 120 K. We focus on the electronic band structure in the vicinity of the Γ point, in particular we measure a $k_y$ cut along the $\Gamma - Y$ direction. Fig. 6a shows the ARPES intensity acquired with the 6.2 eV laser probe. The $x$ axis indicates the emission angle along the $\Gamma - Y$, i.e., here $k_x = 0$. This map was acquired before the arrival of the pump pulse (t = -660 fs). It represents the band structure when probed about 4 $\mu s$ after excitation (the repetition rate of the laser source being 250 kHz). In this way, the electronic band structure approximates the equilibrium one; however, it takes into account average heating effects introduced by pump pulse. Fig. 6b shows the photoemission intensity as a function of $E - E_F$, integrated in a ±1 degree window (dashed white rectangle in Fig. 6a) centered at $k_y = 0$ (Angle=0). Three main features are found: at ~50 meV, ~170 meV, ~360 meV below the Fermi level ($E_F$); they are marked by solid black ticks. These features are in good agreement with both band structure calculations (see Supplementary Information of Ref. [41]) and high-resolution ARPES (the same $k_x = 0$, $k_y \parallel \Gamma Y$ cut is shown in Fig. 1c, as acquired by $h\nu = 68$ eV photons). In the latter case, similar structures - despite the largely different photon energy used - are found. In particular, each of the three structures correspond to a number of hole-like bands that can only be distinguished by high-resolution ARPES. Fig. 6c shows the result of the laser excitation on the electronic band structure via a differential ARPES map (blue color: negative intensity; red color: positive intensity; white color: no intensity change). This map is obtained by subtracting the ARPES map reported in Fig. 6a from the ARPES intensity collected at t = +330 fs (not shown). We chose the pump-probe delay t = +330 fs, for at this delay the photoinduced effect is maximal. Similar to Fig. 6b, Fig. T1d shows the differential ARPES intensity as a function of $E - E_F$, integrated over a ±1 degree window centered about $k_y = 0$, as indicated by the dashed white rectangle in Fig. 6c. The differential ARPES intensity is positive only above $E_F$, indicating that the absorption of the pump pulse heats the sample, leading to broadened band structure features. This effect can arise both by an 'heated' Fermi-Dirac distribution (although this band is below the Fermi energy at the temperature considered) or by an intrinsic broadening of the band at $E - E_F = -30$ meV. We exclude that we are populating electronic states above the Fermi level, since no bands above $E_F$ are expected in this $k_x = 0$, $k_y \sim 0$ region. Below $E_F$, the intensity modification is solely and largely negative. The maximal modification is found in coincidence with the band located 30 meV below $E_F$. We can argue that this band is strongly depleted (and possibly broadened, as argued above) since the energy-position of the maximal modification matches the equilibrium position of the band. Note that a band shift would not be compatible with this finding. The depletion, as induced by excitation with a fluence of 80 $\mu J/cm^2$, is of the order 18%, which is a remarkably large amount. The structure at $E - E_F = -170$ meV shows a similar behavior, although the intensity of the effect is reduced to a few % modification. At variance, the structure at $E - E_F = -360$ meV shows a different response, and possibly shifts to higher binding energies after excitation. This can be argued by the fact that the maximal photoemission intensity variation is peaked at a binding energy different (larger) than the one found at equilibrium (the tick positions are the same as in Fig. 6b). Fig. 7a shows the photoemission energy distribution curves collected at four selected pump-probe delays (and integrated in a window ±1 degree about $k_y = 0$): -660 fs (black line), same as reported in Fig. 6b; +330 fs (red line); +1.3 ps (green line); +4.5 ps (blue line). We can conclude that within 4.5 ps from excitation, the band structure has recovered its equilibrium condition, as indicated by the very good overlap of the energy distribution curves collected at -660 fs and +4.5 ps. After 1.3 ps from excitation, a partial depletion of the $E - E_F = -30$ meV structure is still present. Detailed information about the relaxation timescales have been obtained by analyzing the time evolution of the photoemission intensity in selected energy-momentum windows. We consider four regions, indicated as colored boxes overlaid on Fig. 6c, that have been chosen in order to extract possible band-resolved timescales. The photoemission intensity extracted and averaged from such boxes is reported in Fig. 7b as a function of the pump-probe delay t. The four curves have been analyzed by fitting to them a function composed by one exponential decay with time constant $\tau_j$ ($j = A, B, C, D$) and a positive background, both convoluted with a Gaussian function representing the time resolution of the experiment (pump-probe cross correlation). In the present experiment, the time-resolution is 120 fs FWHM. The constant background at positive delays aims to account for the heat diffusion out of the excited volume. This is a process with nanosecond timescale, and can be safely approximated by a constant in the measured delay range. From the fitting routine (the results of the fit are represented by thin black solid lines) we extract $\tau_A = 1090 \pm 30$ fs, $\tau_B = 970 \pm 30$ fs, $\tau_C = 650 \pm 20$ fs, $\tau_D = 1020 \pm 30$ fs. These values are in good agreement with the findings by TR-ARPES experiments reported in [88]. The relaxation time of ~1 ps also agrees with (momentum-integrated) results from time-resolved optical spectroscopy experiments [89, 90], and is associated to electron-phonon scattering events. From this timescale, the electron-phonon coupling constant was determined [89]. Note that the electron dynamics measured by TR-ARPES do not show the coherent-phonon features revealed by time-resolved reflectivity experiments [89, 90].

By analyzing the non-equilibrium band structure of WTe$_2$ and its ultrafast time-evolution, we can learn important information about its physical properties. In particular, the hole-like bands measured here display a strong depletion which is well beyond what is expected from a simple average heating of the material. In particular, the structure at $E - E_F = -30$ meV is depleted by about 18% after photoexcitation with a moderate fluence. This fact indicates that the possibility to photoinduce a fast imbalance of the electron/hole density ratio is realistic. Obviously, a quantitative estimation of this effect would require the investigation of the time-resolved electronic structure over the full Brillouin zone of WTe$_2$, in particular considering the electron and hole pockets crossing the Fermi level, which are located at $k_x = 0.2 - 0.3$ Å$^{-1}$. The possibility to optically control on ultrafast timescales (less than one ps) the electron-to-hole density ratio, which is considered as the prerequisite for a large and non-saturating magnetoresistance, will have deep implications for the possible optical control of the magnetoresistance on timescales faster than what is currently allowed by electrical means.

## I. Surface distortion and purity of the crystal:

As a last step of refinement of our review on the properties of WTe$_2$, we have studied the surface structure of our cleaved single crystals by high resolution STM. We find that the top layer is tilted by 6°. The tilting is independent of samples and bias voltage. We have also simulated the bandstructure by DFT. We find that although the size of electron and hole pocket depends on the tilting, the electron and hole balance is preserved in the bulk after the inclusion of surface distortion. The details of the STM results and the relation between surface distortion and its effect on the electronic structure is discussed elsewhere [70]. We also find that the surface is extremely clean without any significant impurity and defects. It is possible to find large regions without a single defect. The defect density is about one underlying defect per 3,000 atoms, corresponding to about 9×10$^{11}$ defects cm$^{-2}$. This is important in the context of excitonic dielectric in the Abrikosov sense [91, 92]. Abrikosov proposed that in the quantum limit, where the electron and hole balance is there, the electron and hole pair might form an exciton, quenching the electrical transport. Three main criterion of an excitonic dielectric, i.e., (i) perfect electron-hole balance, (ii) small overlap of valence and conduction bands at Fermi energy [93], and (iii) extremely high purity i.e., low density of impurities and defects [70], seem to be satisfied in this material.

## IV. Summary

We have reviewed the electronic properties of WTe$_2$ by ARPES. We have observed a nearly balanced electron and hole carrier contribution to the Fermi surface, which turns out to be the key reason behind its non-saturating MR behavior as shown by the semiclassical two band model. From our layer dependent first-principles calculations, we find that layer dependence is of extreme importance in the formation of such electron and hole pockets, and their perfect compensation. We find that the electron pocket is characterized by a prominent surface nature, while hole bands are more bulk-like. We find that in order to achieve the perfect electron and hole balance, at least three Te-W-Te layers from the surface must be taken into account, and the balance is indeed maintained in the bulk. We have observed a significant spin polarization of the bands from the Spin resolved ARPES measurements. Such an evidence also confirms that SOC plays a pivotal role in the formation of the electronic structure. Despite being a layered material, WTe$_2$ possesses a band dispersion which is not fully two dimensional. We have probed the 3D Fermi surface by soft-X-ray ARPES measurements and verify that indeed there is a cross layer compensation of carriers. We have also discussed various progresses made on the detection of the topological Weyl Fermi arcs on this system. Finally, we have presented our time-resolved ARPES results and discussed the non-equilibrium dynamics of electronic bands.

## V. Outlook

Currently, there are efforts towards controlled growth of large-scale mono- and few-layer TMDCs in order to integrate them in various real life device applications, which includes field effect transistors, optoelectronic devices, and various sensors applications. Extremely large magnetoresistance of WTe$_2$ is suitable for a magnetic sensor. Note that WTe$_2$ prepared from the ordinary purity W (3N) and Te (4N) exhibit the MR behavior, while to achieve a sizable MR in pure Bi one needs extremely high atomic purity of 99.9995%. WTe$_2$ belongs to the vast family of TMD compounds, putting forward the possibility of chemically doping the material and engineer its properties. WTe$_2$ can be easily exfoliated, and turns out to be suitable for various nanostructured device applications, similar to the ones based on MoS$_2$ [94]. A monolayer of WTe$_2$ is predicted to be a topological insulator [95-100], while it hosts type-II Weyl Fermion states in its bulk [41]. A promising route for future investigations lies in the study of the evolution from one topological configuration to the other as a function of the increasing number of layers. The absence of bulk conduction in monolayer WTe$_2$ could also be related to the excitonic dielectric physics, which albeit requires further experimental verification.

## VI. Acknowledgments

P.K.D. and M.B.H.B. would like to acknowledge the Singapore Synchrotron Light Source (SSLS), which is a National Research Infrastructure under the National Research Foundation Singapore. D.D.S, G.S. and R. T. acknowledge the German Research Foundation (DFG-SFB 1170 Tocotronics), the ERC-StG-336012-Thomale-TOPOLECTRICS and the Gauss Centre for Supercomputing e.V. (www.gauss-centre.eu) for funding this project by providing computing time on the GCS Supercomputer Super-MUC at Leibniz Supercomputing Centre (www.lrz.de). The work at CNR-IOM and CNR-SPIN was performed within the framework of the nanoscience foundry and fine analysis (NFFA-MIUR Italy) project. J. A. K. was supported by the Swiss National Science Foundation (SNF-Grant No. 200021-165910). Part of this work was performed under the approval of Proposal Assessing Committee of HiSOR (Proposal No. 14A24).

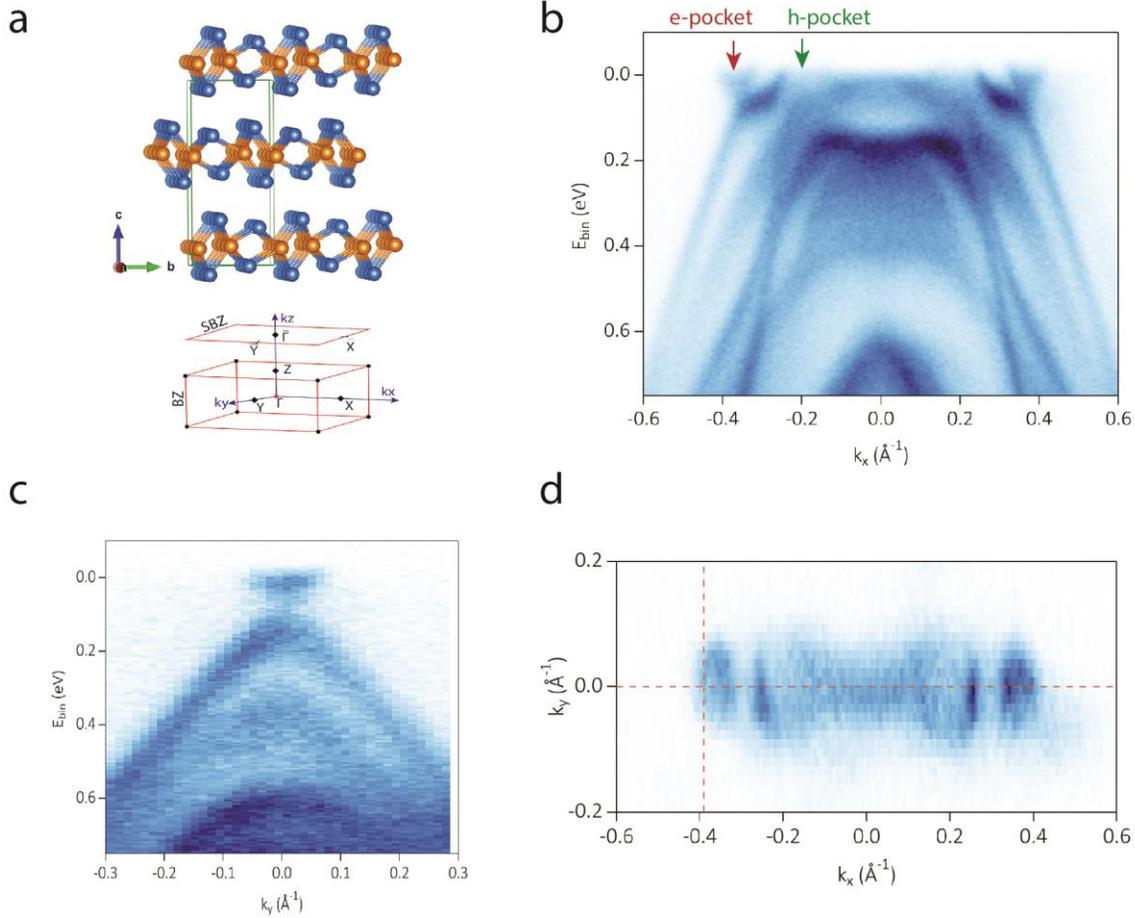

**Fig. 1: Electronic structure of WTe$_2$ measured by ARPES.** (a) Crystal structure of WTe$_2$. A layer of W atoms is sandwiched between two Te layers. Then each di-chalcogenide layers are stacked along the crystallographic *c*-direction and bonded by weak van der Waals coupling. Bottom of the panel a) shows the Brillouin zone and surface Brillouin zone in reciprocal space. (b) band dispersion of WTe$_2$ measured by high resolution ARPES along the $X - \Gamma - X$ direction which is the direction of the Te-W-Te chains. Well defined electron and hole pockets are observed on both sides of the Brillouin zone center ($\Gamma$) as marked by the red and green arrows (shown only on one side), respectively. Panel (c) shows the band dispersion along the $Y - \Gamma - Y$ direction. (d) the Fermi surface map is shown with the dotted lines marking the band cuts shown in panel b and c, respectively.

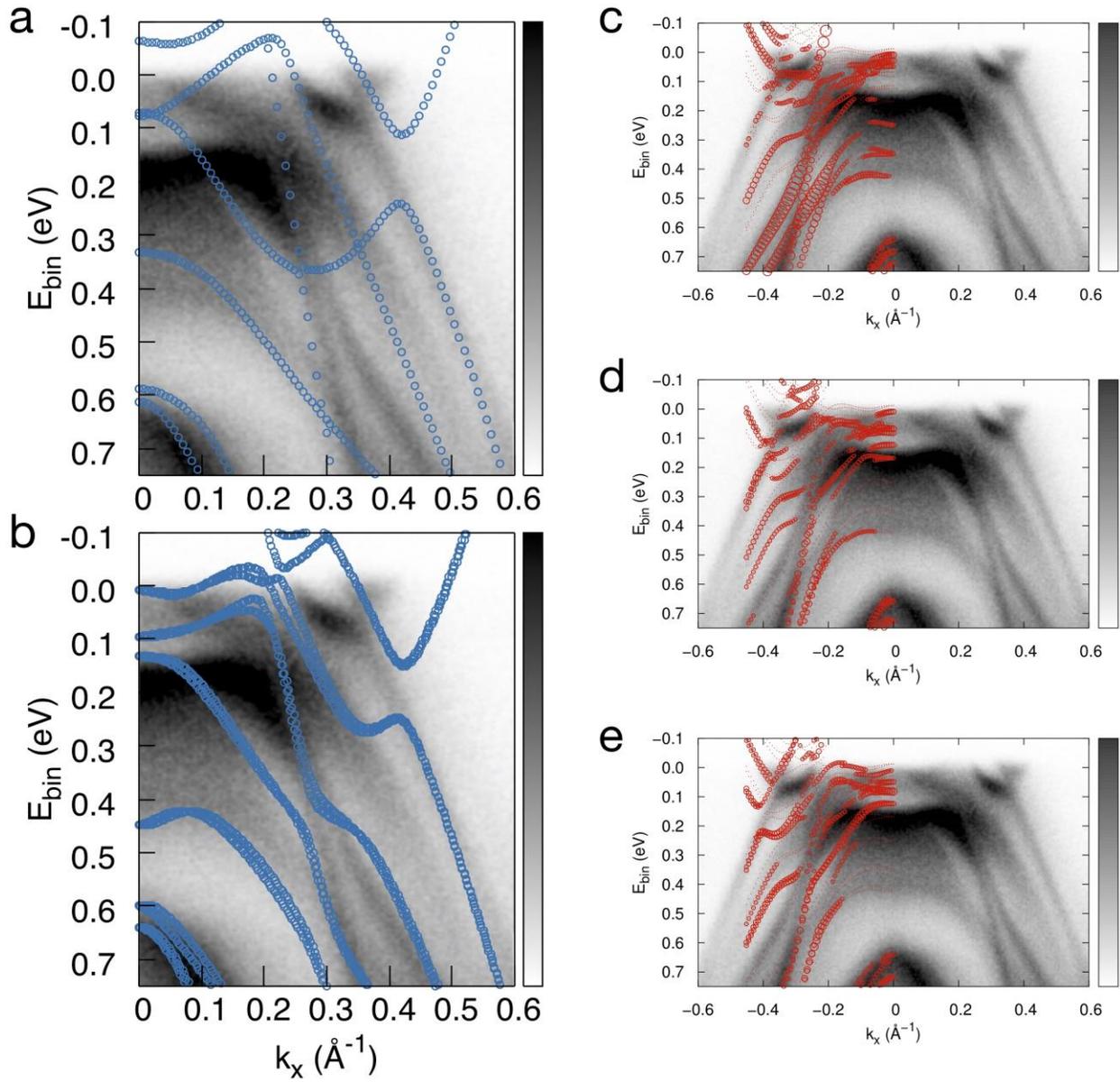

**Fig. 2:** (a-b) Bulk electronic band structure (blue symbols) as calculated without and with the inclusion of SOC, respectively, superimposed to the UV-ARPES measurements ($h\nu$ = 68 eV, T = 77 K) along the $k_x$ direction, i.e. along the W zigzag chains. (c-e) Band structure (red symbols) of a WTe$_2$ finite slab projected onto the 1$^{th}$, 2$^{nd}$ and 3$^{rd}$ WTe$_2$ plane, respectively, superimposed to the same UV-ARPES spectrum as in panels (a-b).

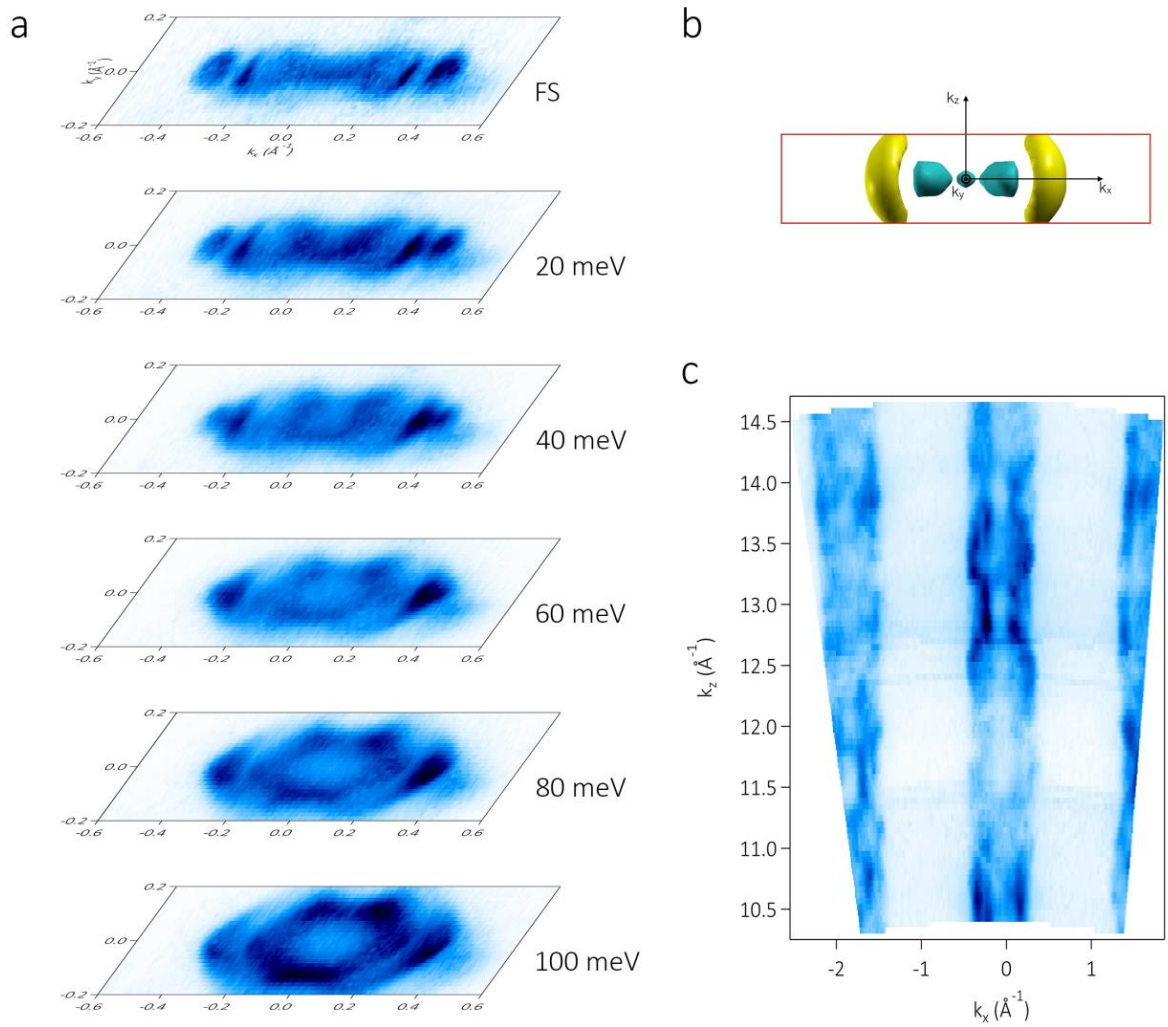

**Fig. 3: Details of the Fermi volume.** (a) Iso-energy $k_x$ vs $k_y$ cuts at the Fermi energy and each 20 meV below down to 100 meV binding energy. (b) simulated $k_x$ vs $k_z$ dispersion, and (c) $k_x$ vs $k_z$ Fermi surface as measured by soft-X-ray ARPES by changing the probe energy from 400 eV to 800 eV.

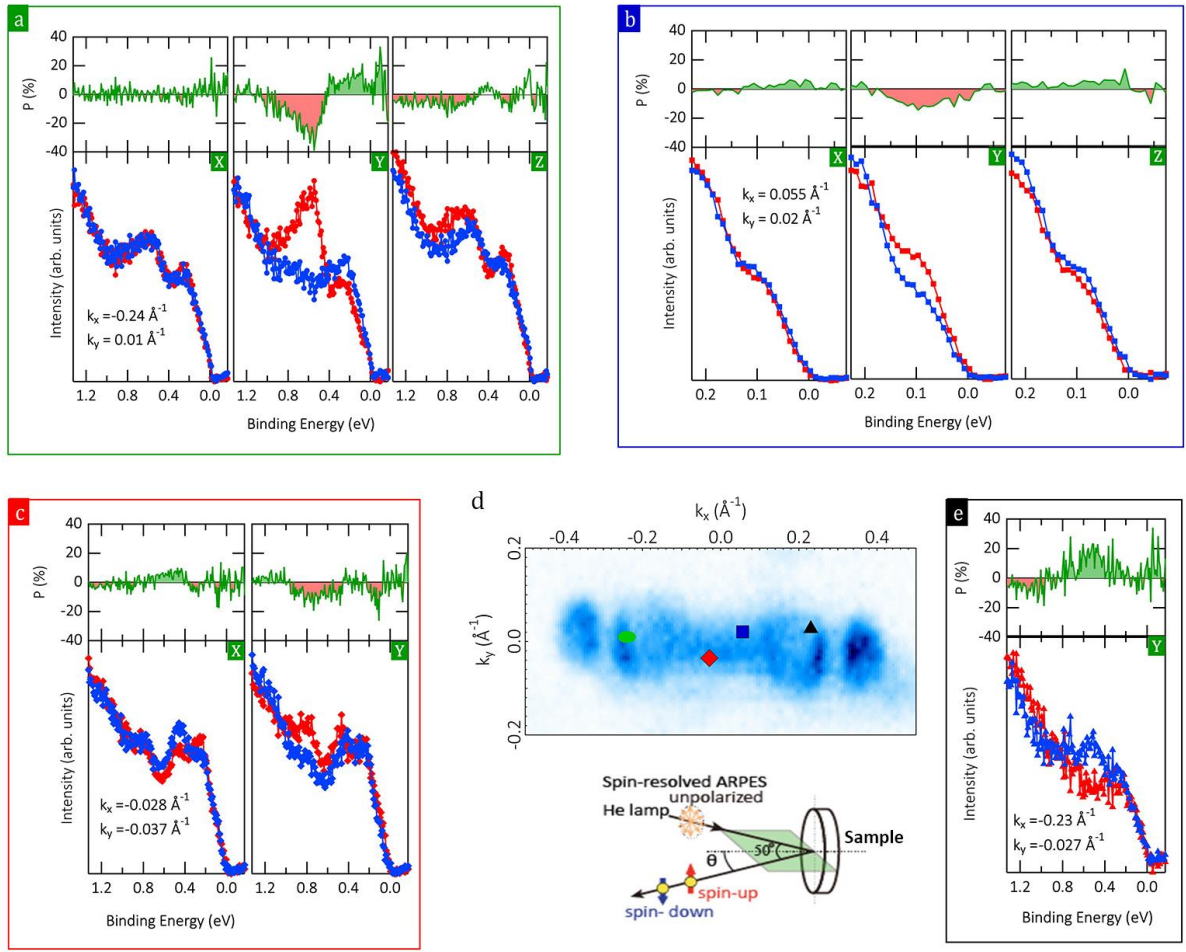

**Fig. 4: Spin texture of WTe$_2$.** The spin polarization of WTe$_2$ bands were probed at four different *k*-points by a VLEED based spin polarimeter. (a) presents the spin resolved EDCs measured at near the edge of the hole pocket as shown in the Fermi surface map in panel d by the green circle. Blue and red colored spectra corresponds to the up and down spin orientations, respectively. We see a significant spin polarization of 40% along the y-direction at higher binding energy values. Although we observed smaller spin polarization value at the Fermi edge, it is non-negligible at some *k*-positions as shown in panel (b). Panel (c) and (e) shows the spin polarized EDCs measured at positions as indicated by red diamond (near Γ point but away from Γ − X lines) and black triangle (edge of hole pocket opposite of green dot position) over the Fermi surface in panel d. The bottom panel of d shows the geometry of spin resolved ARPES setup.

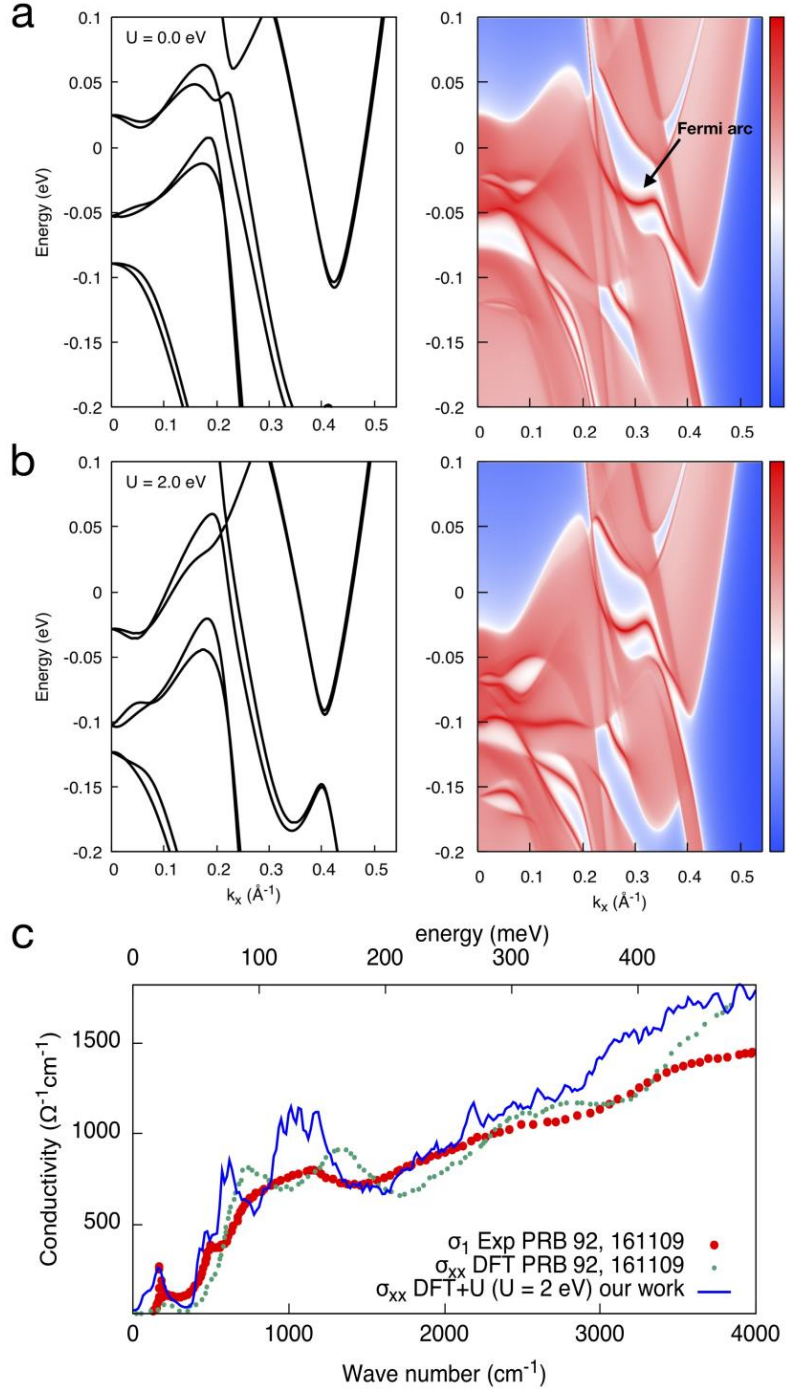

**Fig. 5:** (a-b) DFT+U band structure along the $k_x$ direction (left panels) and corresponding theoretical spectral function for the (001) surface (right panel) as calculated for (a) U = 0.0 eV and (b) U = 2.0 eV. In the spectral function shown in panel (a) the black arrow highlights the Fermi arc which connects the hole and electron pockets in a type-II Weyl semimetal. (c) The experimental optical conductivity (red dots) is compared with the theoretical optical conductivity calculated by setting U = 0.0 eV (green dots) and U = 2.0 eV (blue solid line).

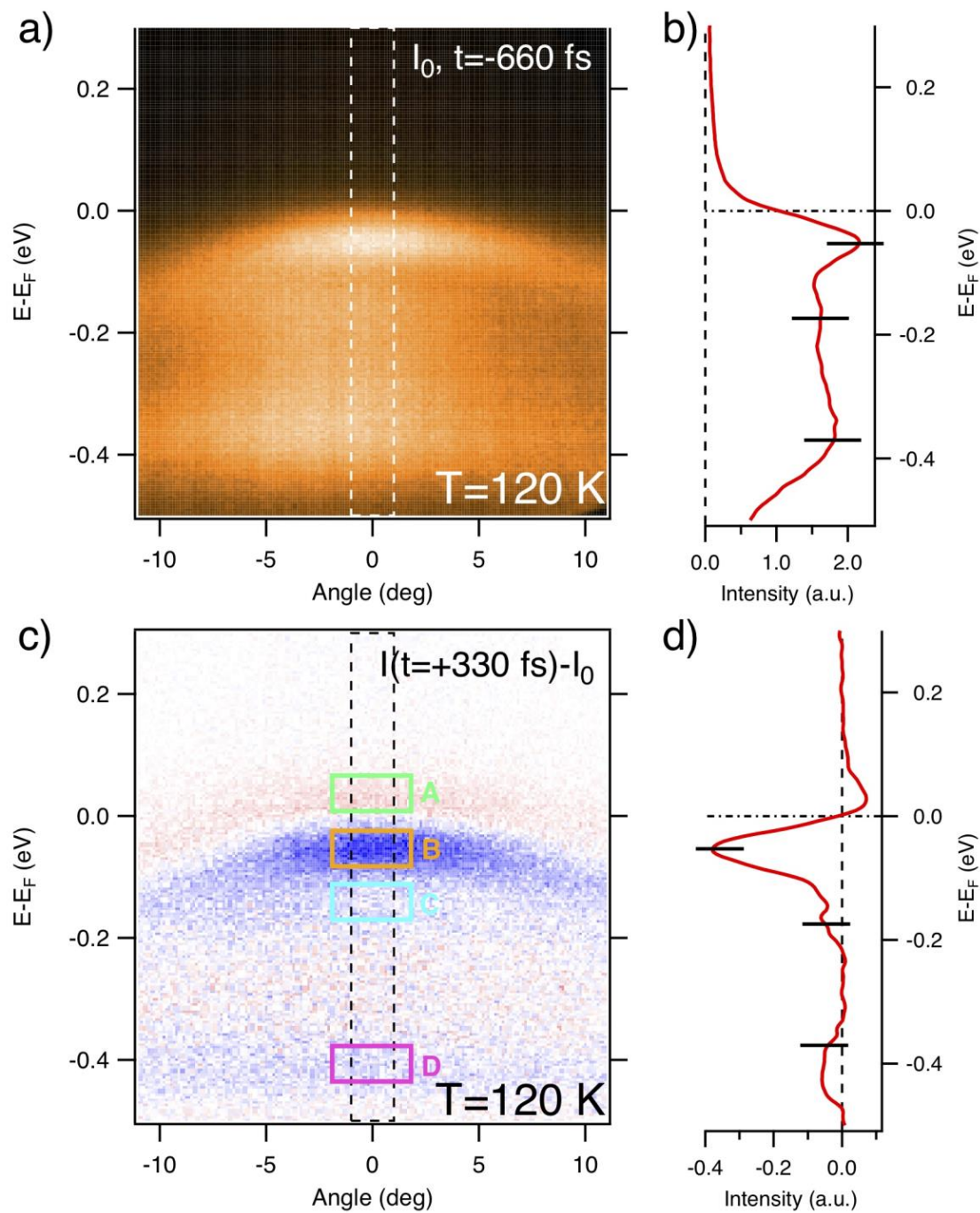

**Fig. 6:** Non-Equilibrium Electronic Structure by TR-ARPES Experiments. (a) The electronic band structure along $\Gamma - Y$ direction ($k_x$=0), measured at $h\nu$=6.2 eV and T=120 K. The map has been acquired at t=-660 fs. (b) Energy-distribution curve integrated in a ±1 degree window centered about $k_y$=0 (white dashed rectangle in panel a). Three main structures, in agreement with findings

reported in Fig. 1c, are found at E-E$_F$= ~30 meV, ~170 meV, ~360 meV. (c) Differential ARPES map acquired at t=+300 fs (see text). Red color indicates an increase in photoemission intensity, blue color indicates a photoemission intensity decrease. (d) Differential energy distribution curve extracted from the white dashed rectangle indicated in panel c. The photoemission intensity increases above the Fermi level, while decreases below E$_F$, with a marked depletion centered at E-E$_F$=~30 meV.

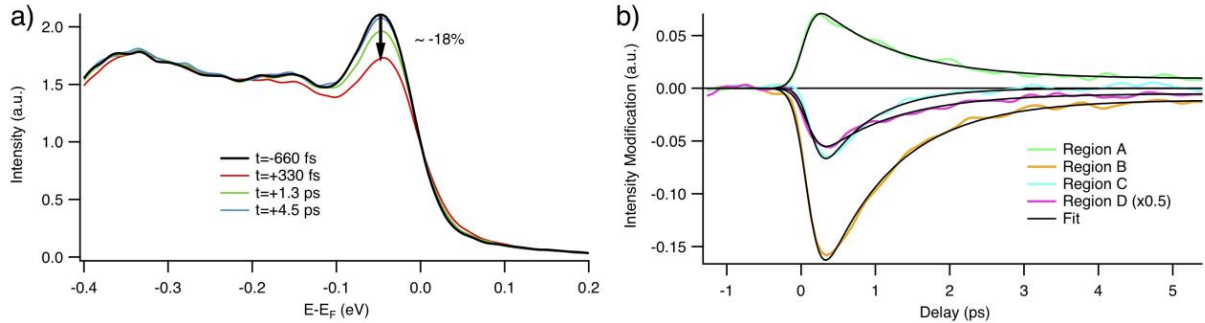

**Fig. 7:** Non-equilibrium dynamics. (a) Energy distribution curves collected at t=-660 fs, t=+330 fs, t=+1.3 ps, t=+4.5 ps. (b) Electron dynamics extracted at selected energy-momentum windows, as indicated in Fig. 6c. The solid black lines are fit to the data of an exponentially decaying function.